\documentclass[aps, preprint, groupedaddress, floatfix, notitlepage]{revtex4-2}
\usepackage{epsfig}
\usepackage{graphicx}
\usepackage{amsmath}
\usepackage{bbold}
\usepackage{tabularx, booktabs}
\usepackage[normalem]{ulem}
\usepackage{xcolor}
\usepackage{multirow}
\usepackage{dutchcal} % curly characters
\usepackage[T1]{fontenc}
\usepackage{longtable}
\usepackage{appendix}

\usepackage{float}
\usepackage[caption = false]{subfig}
\newcolumntype{Y}{>{\centering\arraybackslash}X}
\newcolumntype{Z}{>{\hsize=1.1\hsize\centering\arraybackslash}X}

\begin{document}

\title{
Manipulation of magnetic anisotropy of 2D magnetized graphene by ferroelectric In$_2$Se$_3$ 
}

% \date{\today}
\author{Rui-Qi Wang$^{1,2}$}
\author{Tian-Min Lei$^{1}$}
\email{leitianmin@163.com}
\author{Yue-Wen Fang$^{3,4}$}
\email{fyuewen@gmail.com}

% \affiliation{$^1$Fisika Aplikatua Saila, Gipuzkoako Ingeniaritza Eskola, University of the Basque Country (UPV/EHU), San Sebasti\'{a}n, Spain\\
\affiliation{$^1$ School of Advanced Materials and Nanotechnology, XiDian University, Xi’an 710126, China \\
$^2$ School of Electronic Engineering, Xi’an Aeronautical Institute, Xi’an 710077, China \\
$^3$ Fisika Aplikatua Saila, Gipuzkoako Ingeniaritza Eskola, University of the Basque Country (UPV/EHU), Europa Plaza 1, 20018 Donostia/San Sebasti{\'a}n, Spain \\
$^4$  Centro de F{\'i}sica de Materiales (CSIC-UPV/EHU), Manuel de Lardizabal Pasealekua 5, 20018 Donostia/San Sebasti{\'a}n, Spain \\
}
\begin{abstract}
\vspace{1cm}
The capacity to externally manipulate magnetic properties is highly desired from both fundamental and technological perspectives, particularly in the development of magnetoelectronics and spintronics devices. Here, using first-principles calculations, we have demonstrated the ability of controlling the magnetism of magnetized graphene monolayers by interfacing them with a two-dimensional ferroelectric material. When the 3$d$ transition metal (TM) is adsorbed on the graphene monolayer, its magnetization easy axis can be flipped from in-plane to out-of-plane by the ferroelectric polarization reversal of In$_2$Se$_3$, and the magnetocrystalline anisotropy energy (MAE) can be high to -0.692 meV/atom when adopting the Fe atom at bridge site with downward polarization. This may be a universal method since the 3$d$ TM-adsorbed graphene has a very small MAE, which can be easily manipulated by the ferroelectric polarization. As a result, the inherent mechanism is analyzed by second variation method.
\end{abstract}
\maketitle
\newpage

\section{Introduction}

One fundamental concept in condensed matter physics is that most storage devices are based on ferromagnetism. A large magnetic anisotropy (MA) can act as the magnetization reversal barrier to prevent thermal disturbance, which is the key to realizing the monoatomic information storage. In order to read and write the information of an individual magnetic atom, ferroelectric-controlled magnetism has become a popular way. Both the states with polarization upward ($P\uparrow$) and polarization downward ($P\downarrow$) in the ferroelectric material are stable, which can be switched between each other via an external electric field. This makes the ferroelectric material a bistable switch and has a wide range of application prospects in electric-controlled magnetism. 

Ferroelectric-controlled magnetism refers to the control of magnetic properties such as magnetic moments, magnetic anisotropies, magnetic orders, or magnetic resistance through ferroelectrics, sometimes with auxiliary conditions like stress, charge injection, electric field, magnetic field, and temperature, etc. The researchers have successively realized the ferroelectric-control of ferromagnetic-antiferromagnetic transition in two-dimensional ferroelectric/ferromagnetic heterojunctions such as In$_2$Se$_3$/CrI$_3$[1-2], Sc$_2$Co$_2$/CrI$_3$[3], Sc$_2$Co$_2$/NiI$_2$[4], and In$_2$Se$_3$/FeI$_2$[5]. Ferroelectric-control of the easy axis of magnetization is achieved in two-dimensional ferroelectric/ferromagnetic heterojunctions such as In$_2$Se$_3$/NiI$_2$[6], In$_2$Se$_3$/FeBrI[7],  In$_2$Se$_3$/CrI$_3$/HfN2[8], Sc$_2$Co$_2$/Cr$_2$Ge$_2$Te$_6$[9-10], and In$_2$Se$_3$/3$d$ transition-metal atoms[11]. The latter ones are of great reference significance for the study of this paper.

In order to successfully realize the magnetization easy axis flip in the ferroelectric/ferromagnetic interface, the initial MAE of the ferromagnet needs to be small, while the polarization of the ferroelectric needs to be large. Furthermore, the magnetoelectric coupling effect should be strong.

As a new 2D ferroelectrics, $\alpha$-In$_2$Se$_3$ has stable spontaneous in-plane and out-of-plane ferroelectric polarizations at room temperature, and can still maintain them under the limited thickness of a single layer[12]. Quantitative experimental results show that an ultrathin $\alpha$-In$_2$Se$_3$ layer owns a spontaneous polarization $P_{\rm S}$ = 0.92 ${\rm \mu}$C/cm$^2$ under the external electric field of 5$\times$10$^5$ V/cm[13]. It has recently been theoretically shown that switching the direction of electrical polarization of $\alpha$-In$_2$Se$_3$ could affect the magnetization direction of its adlayers[6, 8, 11].

Another key technology is to select a ferromagnet with a small MAE. In recent years, a number of two-dimensional ferromagnetic materials have emerged. Cr$_2$Ge$_2$Te$_6$ has been experimentally found to have intrinsic magnetism[14], which can be controlled by adjacent ferroelectricity[9]. CrI$_3$ is proved to be an Ising ferromagnet with an out-of-plane spin orientation by magneto-optical Kerr effect microscopy[15]. Researchers also find strong anisotropy and magnetostriction in 2D Stoner ferromagnet Fe$_3$GeTe$_2$[16]. Furthermore, the quantum anomalous valley Hall effect has been found in both ferromagnetic monolayer RuClBr[17] and antiferromagnetic monolayer MoO[18]. Intrinsic Dirac half-metal and quantum anomalous Hall phase have been discovered in the hexagonal Nb$_2$O$_3$ lattice[19].

Instead of using the latest two-dimensional ferromagnets, we use transition metals to adsorb on graphene[20-21] not only for the graphene’s good electrical and mechanical properties[22] but also for the systems’ small MAEs. In other words, the selected lattice must have a very small initial MAE to be easily controlled by the adjacent ferroelectric. In this study, the rhombic lattice of graphene was chosen to adsorb the 3$d$ TM atoms, which granted the graphene with a very small MAE, and then the two-dimensional In$_2$Se$_3$ is utilized to manipulate the magnetization easy axis of the magnetized graphene monolayers. Better than the previous study where MAE reached -0.23 meV/atom at most[23], we show high magnetic crystal anisotropy (up to -0.692 meV/atom) in this study which is essential for being used in new magnetic storage. In other words, by manipulating the occupation of some specific 3$d$ orbitals, we can achieve accurate regulation of the MAE. We find it a universal method to manipulate the easy magnetization axis by the ferroelectric polarization since the magnetized graphene has a very small initial MAE.

\section{Methods}\label{sec:DFT-methods}
Fig. 1 shows the heterojunction of In$_2$Se$_3$/TM-adsorbed graphene, in which the In$_2$Se$_3$ monolayer and the magnetized graphene are adopted to form a heterojunction. We choose a rhombic graphene monolayer with the lattice constant a = 4.26 \AA. The optimized lattice constant of 1$\times$1 In$_2$Se$_3$ is 4.11 \AA. To make up for the small lattice mismatch ($\sim$3.5$\%$), the lattice constant of In$_2$Se$_3$ is enlarged by 3.5$\%$, while the lattice constant of graphene is fixed at the value of 4.26 \AA. To eliminate the interaction between adjoining heterostructures, a 20 \AA~vacuum layer along the $z$-axis is included. Fig. 2 shows the rhombic interfacial shape of the lattice, where the graphene monolayer adsorbs the individual TM atom on three different sites (hollow, bridge, and top), respectively.

%%%%%%%%%%%%%%%%Figure 1 phase diagram
\begin{figure*}[h]
\includegraphics[angle=0,width=0.7\textwidth]{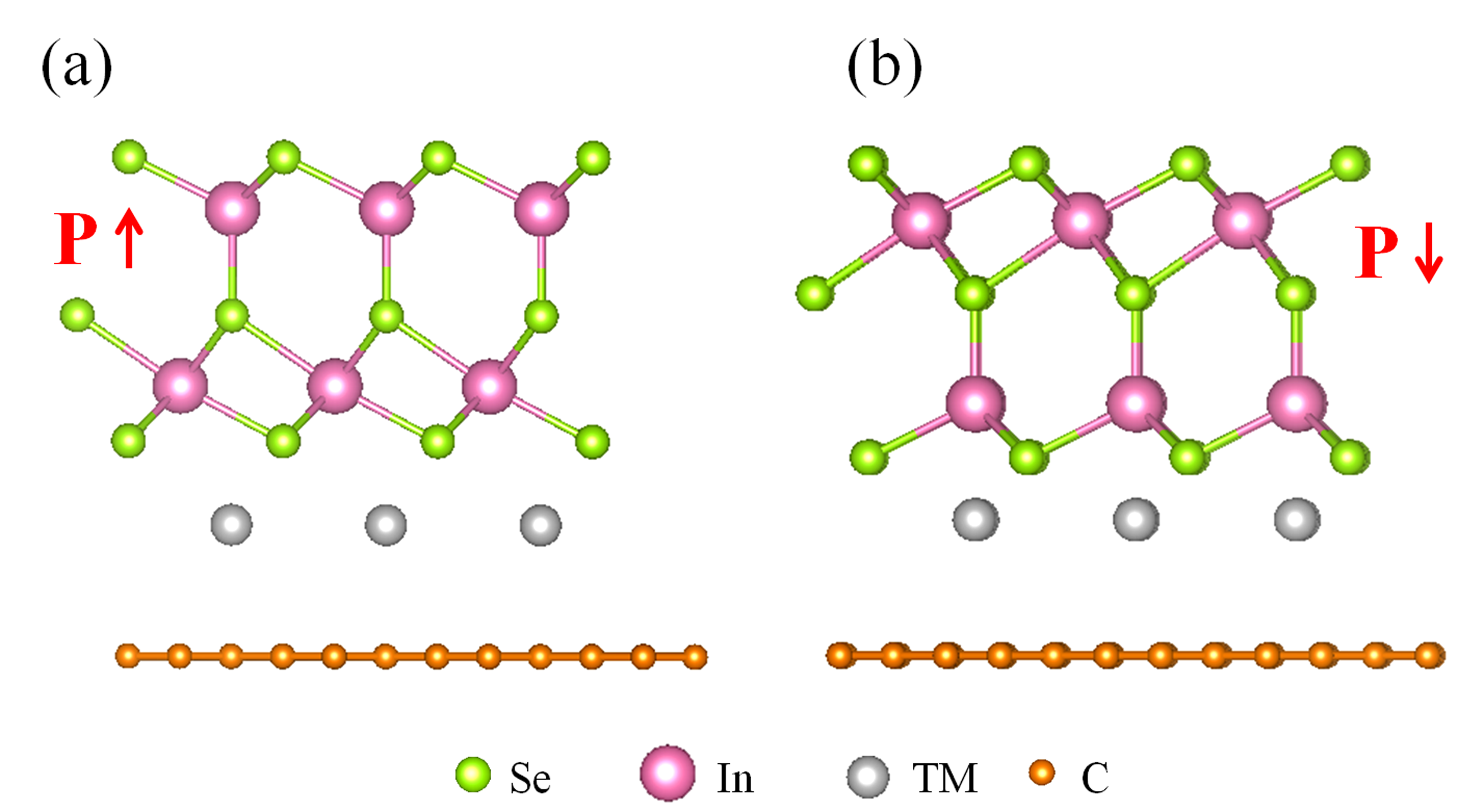}
\caption{\label{fig:1}  
Atomic structures of In$_2$Se$_3$/graphene heterostructures. The Se, In, TM, and C atoms are represented by green, purple, gray, and orange balls, respectively. Polarization of In$_2$Se$_3$ is from the negative Se atom to the positive In atom: (a) the direction of polarization is vertically upward; (b)the direction of polarization is vertically downward.
}
\end{figure*}

%%%%%%%%%%%%%%%%Figure 2 phase diagram
\begin{figure*}[h]
\includegraphics[angle=0,width=0.7\textwidth]{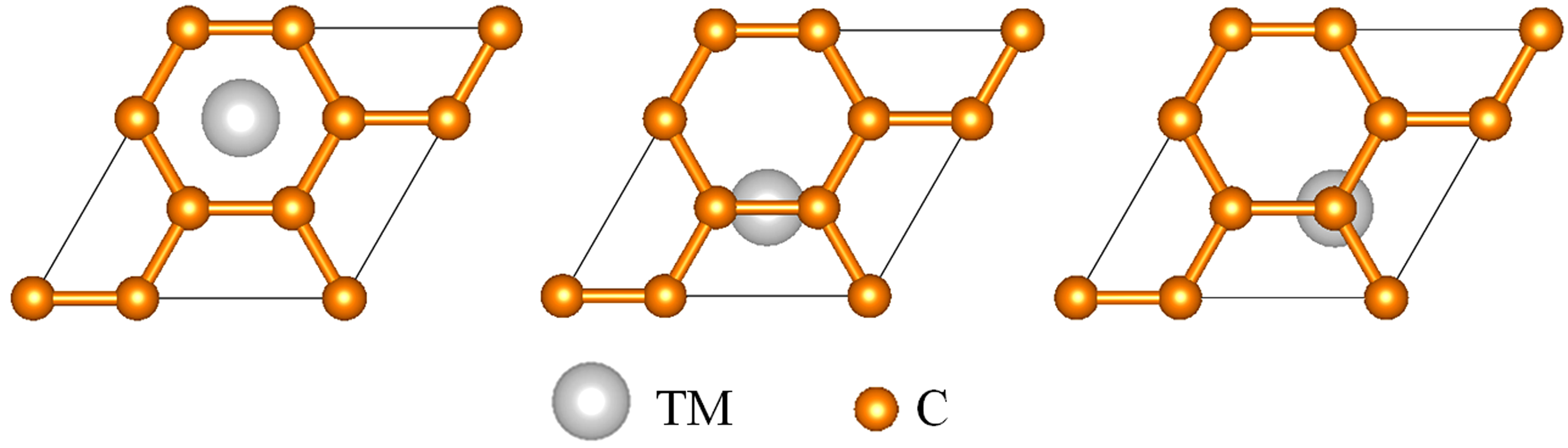}
\caption{\label{fig:2}  
The rhombic interfaces of the heterojunctions. Transition metal atoms are attached to the surfaces of graphene in three different ways (hollow, bridge, and top). In$_2$Se$_3$ has been left out for a better view of the interfaces. 
}
\end{figure*}

The calculations are performed in the Vienna Ab-initio Simulation Package (VASP) based on density-functional theory (DFT). The DFT + $U$ approach[24] is essential for the strongly correlated electronic systems with partially filled 3$d$-shells. In this work, we use $U$ = 2 eV and $J$ = 0 eV to study the magnetism, which are the typical values for 3$d$ TM atoms[25]. Furthermore, the interfacial distances have been optimized by the DFT-D3 method[26]. The generalized gradient approximation (GGA) with the form of Perdew–Burke–Ernzerhof (PBE) is utilized to describe the exchange–correlation functional. A 16$\times$16$\times$1 Monkhorst-Pack k-point grids is good enough to sample the Brillouin zone. The cut-off energy is set to be 480 eV and the criterion for the convergence of total energy is set to 10$^{-6}$ eV. The convergence test of total energy and MAE can be seen from Fig. S1-S6 in Supplementary Material. The dependency between the value of MAE and Ueff has also been tested and it is found that different Ueff values may not change the positive or negative signs of MAE, and only alter the magnitude of MAE slightly (see Fig. S7-S9). Moreover, the screened exchange hybrid density functional by HeydScuseria-Ernzerhof (HSE06) [27] is utilized to check the electronic structure (see Fig. S10).

The following definition represents the binding energy of the TM/graphene system[28]:
\begin{equation}
    E_b = E_{\rm TM/graphene} - E_{\rm TM} - E_{\rm graphene}
\end{equation}
Here $E_{\rm TM/graphene}$, $E_{\rm TM}$, and $E_{\rm graphene}$ are the total energies of the TM/graphene system, the TM atom and the graphene slab, respectively. Table 1 shows the calculated binding energies with 3$d$ TM atoms on graphene monolayer at different sites. These values reveal that the most stable position of Mn, Fe, and Co is the bridge site, while it is the top site for V and Cr atoms, and only the Ni atom is most stable at the hollow site.

\begin{table}
\centering
\caption{Calculated binding energies (in units of eV) for the TM atom-adsorbed graphene monolayers at bridge, hollow, and top sites.}
\begin{tabular}{lccc}
\toprule
TM atom & Bridge & Hollow & Top \\
\midrule
V  & -0.7593 & -0.6025 & -0.8034 \\
Cr & -0.2626 & -0.2386 & -0.2689 \\
Mn & -0.2921 & -0.2918 & -0.2914 \\
Fe & -0.7288 & -0.4607 & -0.6594 \\
Co & -0.3737 & -0.1504 & -0.3689 \\
Ni & -0.7819 & -0.8015 & -0.7776 \\
\bottomrule
\end{tabular}
\end{table}

%%%%%%%%%%%%%%%%Figure 3 phase diagram
\begin{figure*}[h]
\includegraphics[angle=0,width=0.7\textwidth]{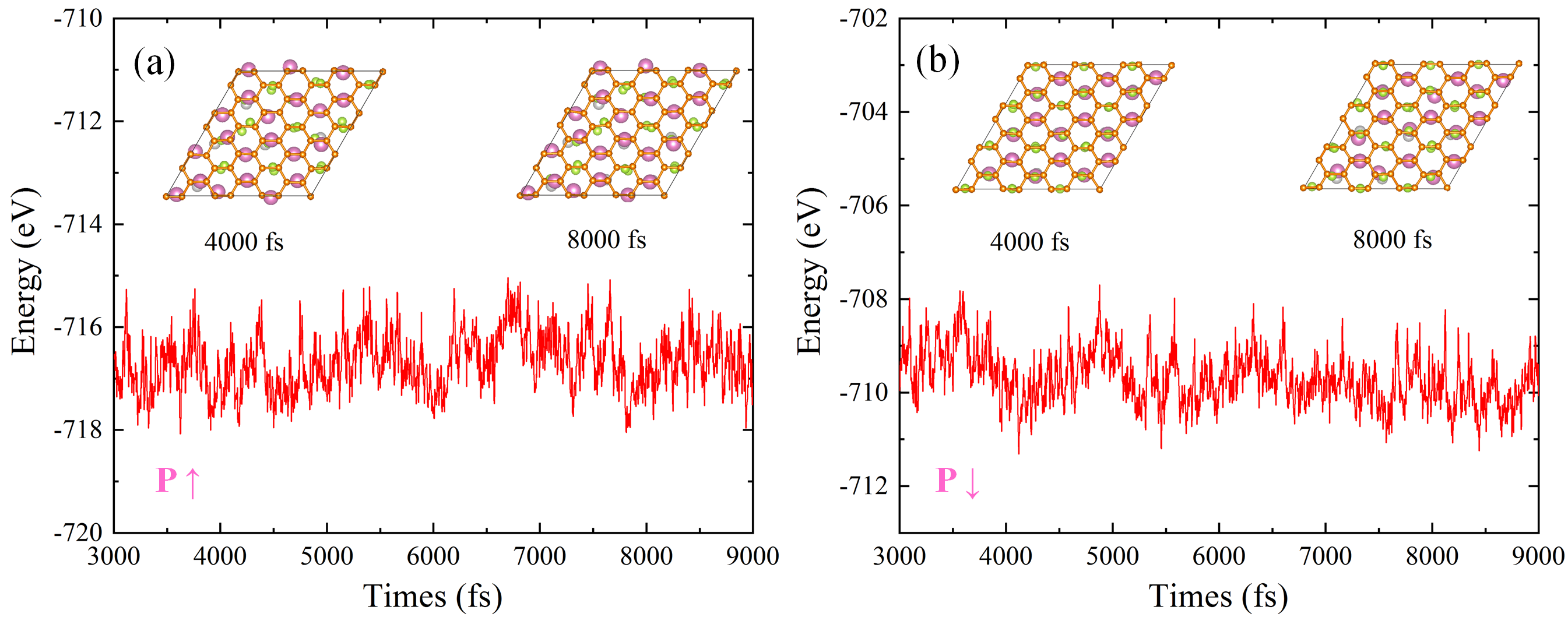}
\caption{\label{fig:3}  
Variation of the energy from 3000 to 9000 fs during AIMD simulations at a temperature of 300 K for the In$_2$Se$_3$/Fe bridge-site adsorbed graphene heterostructures: (a) $P\uparrow$ and (b) $P\downarrow$. The inset illustrates the sliced snapshots of the crystal structures at 4000 and 8000 fs, respectively.
}
\end{figure*}

Through careful calculations, various interfacial configurations of In$_2$Se$_3$ and graphene were checked to consider the magnetoelectric coupling effect. Ultimately, the interfaces in which the TM atoms were exactly aligned with the In-Se bonds along the $z$-axis were chosen since their strong magnetoelectric coupling effect achieved the ferroelectric manipulation of the magnetization easy axis.
The molecular dynamics simulations are performed at a temperature of 300 K with a 3$\times$3 supercell in Fig. 3. The energy fluctuations are small in the time range of 3000–9 000 fs, indicating that the structures are basically mechanically stable.

The MAE is obtained by the difference of total energies when the magnetization directions are along [100] and [001], thus the positive MAE refers to the out-of-plane MA while the negative MAE refers to the in-plane MA. First of all, the charge density is gained through the calculation of self-consistent without the spin-orbit coupling (SOC) effect. Then the noncollinear calculations are carried out for the MAE by reading the above charge density and considering the SOC effect. The convergence of MAE is achieved with a \textbf{k}-point meshes of 30$\times$30$\times$1 (see Fig. S6).

\section{Results \& Discussions}
We adopt the most stable adsorbing types for each transition metal atom adsorbed on graphene. The magnetic moments of 3$d$ TM atoms with the ferroelectric $P\uparrow$ and $P\downarrow$ are shown in Fig. 4 for comparison. It is obvious that most of the magnetic moments are non-zero for TM atoms, except for Ni atoms at the $P\downarrow$ state, where the occupations of spin up and spin down are equal. When the ferroelectric polarization of In$_2$Se$_3$ is turned from $P\uparrow$ to $P\downarrow$, magnetic moments of the V, Fe, Co, and Ni atoms diminish, whereas those of the Cr and Mn atoms increase.

As can be clearly seen in Fig. 5, the MAE of the 3$d$ TM-adsorbed graphene monolayer without ferroelectric polarizations is generally below 0.100 meV/atom and the maximum is only about 0.250 meV, which is the key prerequisite to be easily manipulated by the adjacent ferroelectric polarization as previously mentioned.

%%%%%%%%%%%%%%%%Figure 4 phase diagram
\begin{figure*}[h]
\includegraphics[angle=0,width=0.7\textwidth]{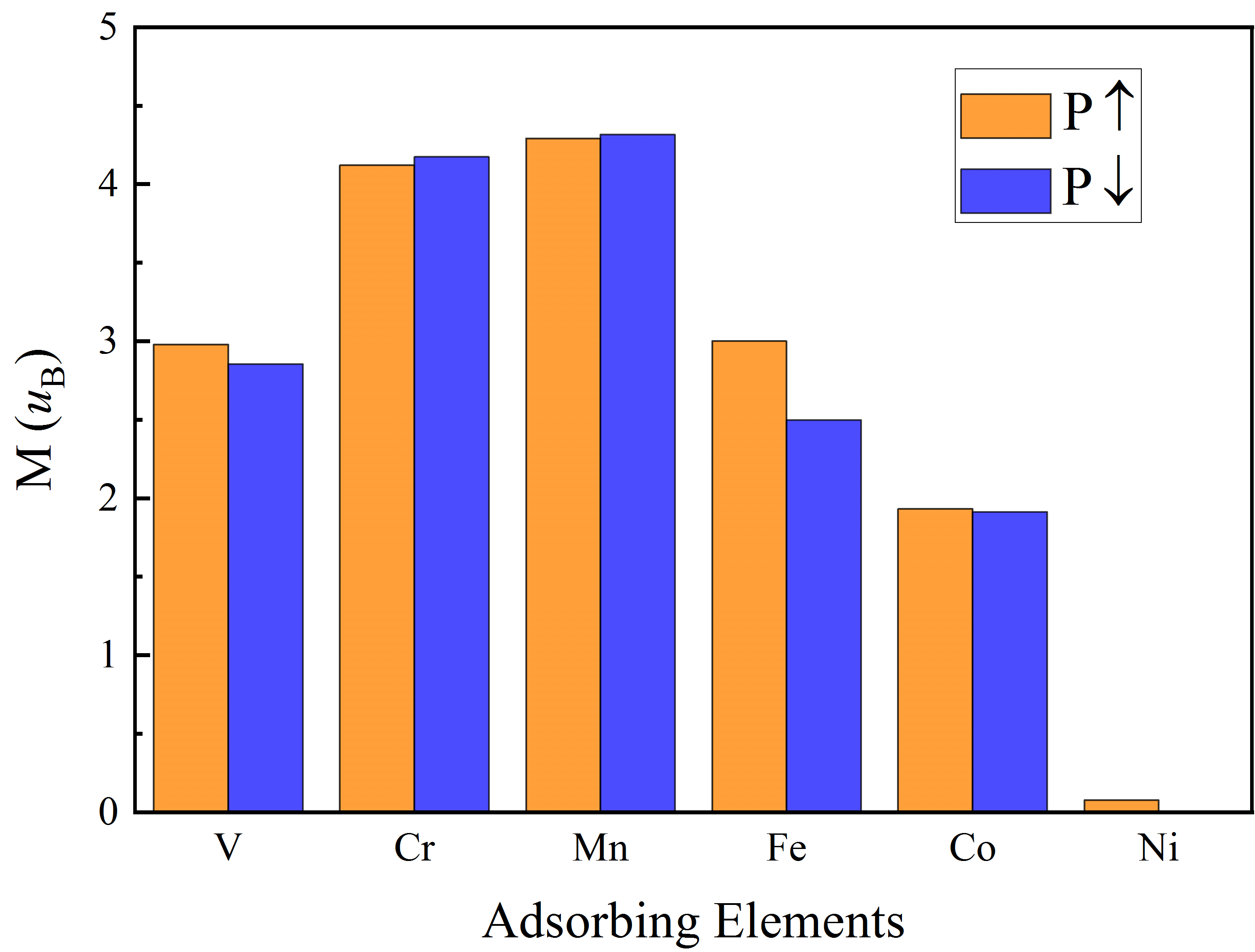}
\caption{\label{fig:4}  
Magnetic moments of TM atoms in the In$_2$Se$_3$/TM adsorbed graphene heterojunctions for the $P\uparrow$ and $P\downarrow$ of In$_2$Se$_3$.
}
\end{figure*}

%%%%%%%%%%%%%%%%Figure 5 phase diagram
\begin{figure*}[!ht]
\includegraphics[angle=0,width=0.7\textwidth]{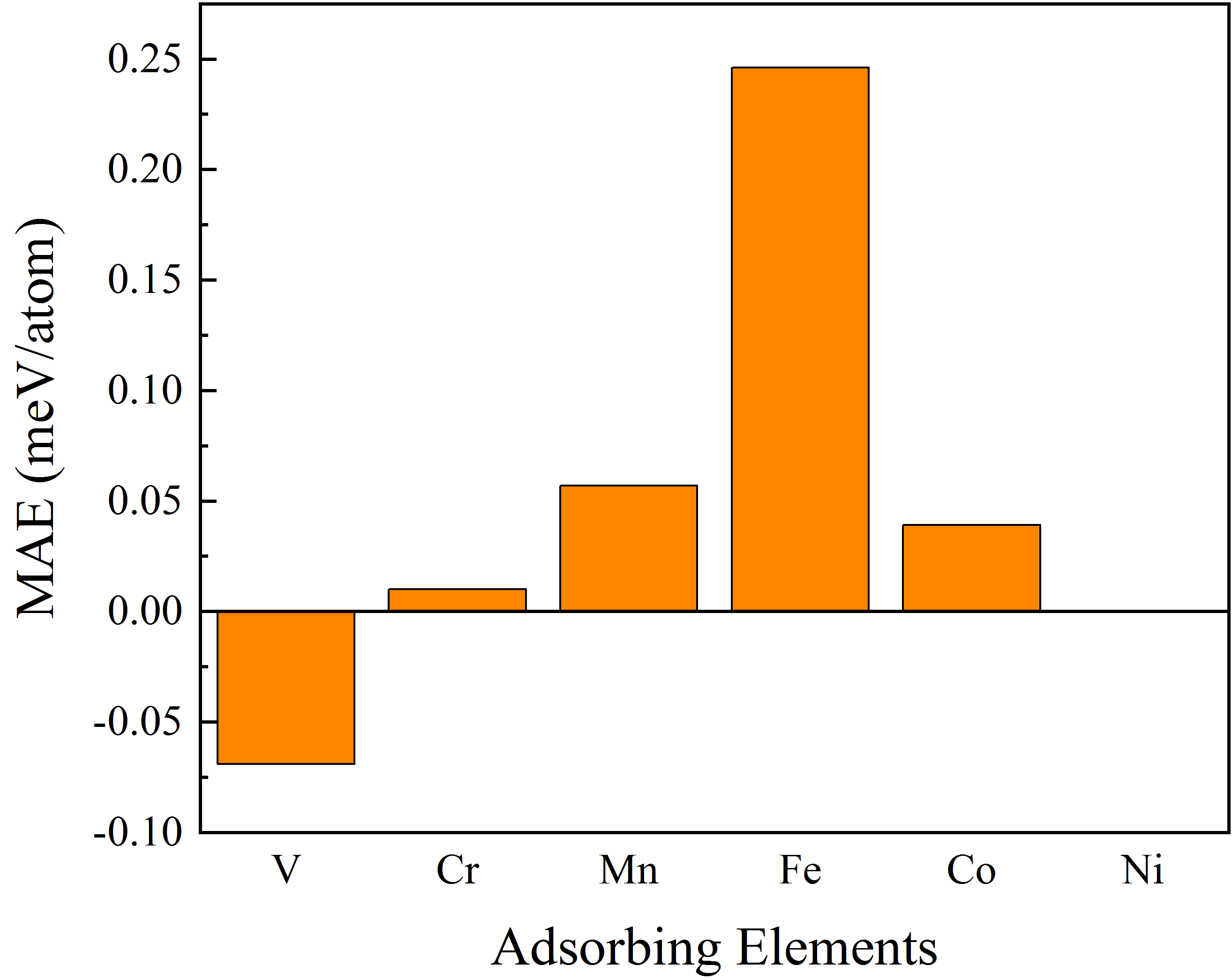}
\caption{\label{fig:5}  
MAE of the TM atom-adsorbed graphene monolayers without ferroelectric polarizations.
}
\end{figure*}

%%%%%%%%%%%%%%%%Figure 6 phase diagram
\begin{figure*}[!ht]
\includegraphics[angle=0,width=0.7\textwidth]{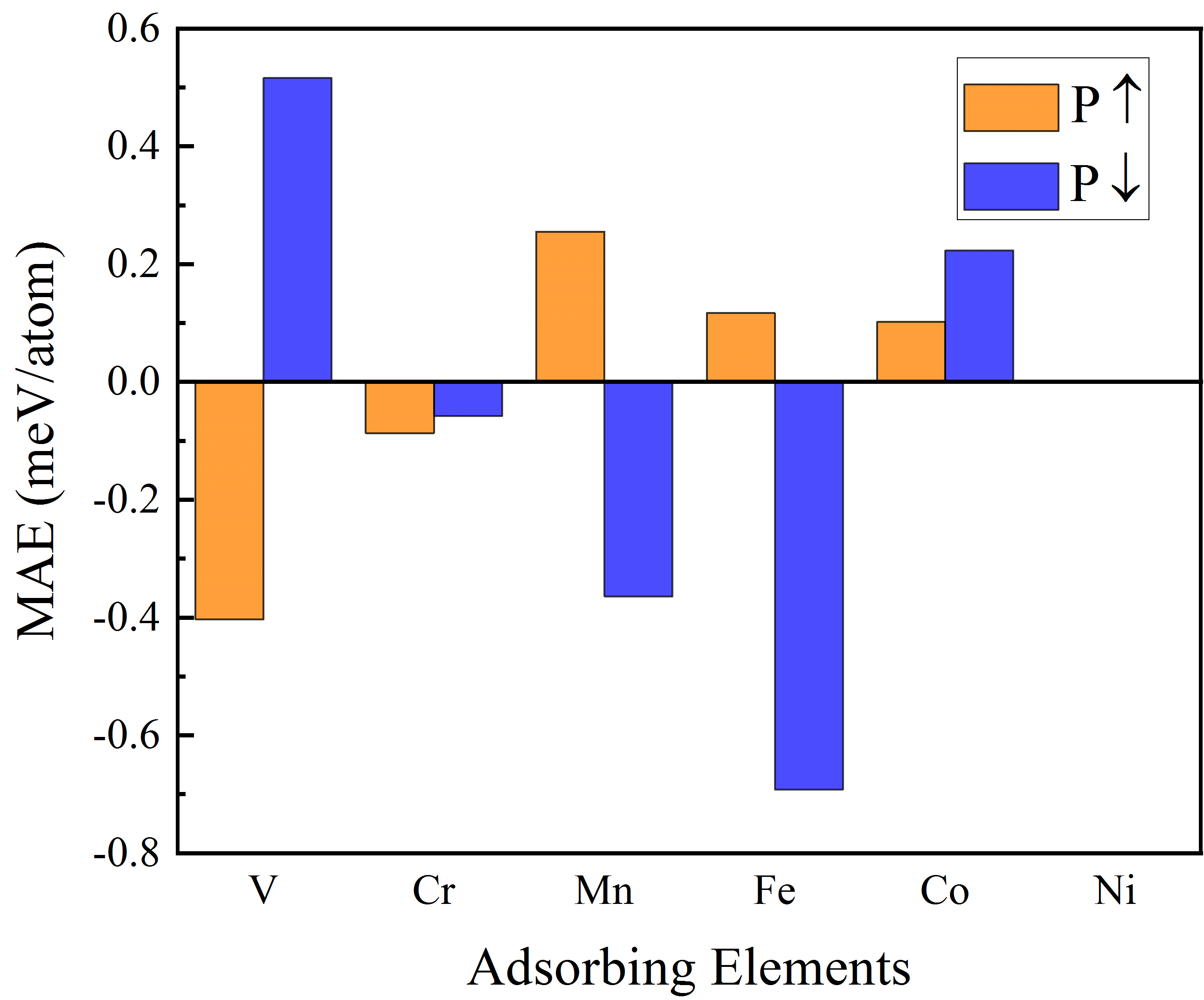}
\caption{\label{fig:6}  
MAE of the In$_2$Se$_3$/TM adsorbed graphene heterojunctions for the $P\uparrow$ and $P\downarrow$ of In$_2$Se$_3$.
}
\end{figure*}

When added by extra ferroelectric polarizations of In$_2$Se$_3$, the MAEs for different TM atoms-adsorbed graphene monolayers are plotted in Fig. 6 for comparison. It is shown that the In$_2$Se$_3$ polarization inversion from $P\uparrow$ and $P\downarrow$ can weaken the in-plane MA at the Cr atom, and strengthen the out-of-plane MA at the Co atom. Whereas, the most exciting finding is that when 3$d$ TM atoms are V, Mn, and Fe, the easy axes of magnetization in these systems all depend on the ferroelectric polarization direction of In$_2$Se$_3$. For Mn and Fe atoms, the MAE is positive in  $P\uparrow$ state but negative in  $P\downarrow$ state. Whereas, the correspondence is reversed for V atoms. As a result, we realized the flip of magnetization easy axis from out-of-plane to in-plane via ferroelectric control. Here we select Fe atoms for further research, since the $P\uparrow$ state has a positive MAE of 0.117 meV/atom, while the $P\downarrow$ state has a negative MAE of -0.692 meV/atom, which is the largest MAE value in our study. It should be noted that the MAE of the Ni atom is zero regardless of whether the ferroelectric is added or not (see Fig. 5 and Fig. 6) since the magnetic moment is almost zero.

Through calculating the Berry Phase[29], the polarization of ferroelectrics is carried out by $P$ = ($P\uparrow$ - $P\downarrow$ ) / 2 where $P\uparrow$ and $P\downarrow$ are the ferroelectric polarization along $z$ and -$z$ axis. Well consistent with the predecessors' study[30], the polarization of monolayer In$_2$Se$_3$ is calculated to be 0.96 ${\rm \mu}$C/cm$^2$. Furthermore, If $P$ = +0.96  ${\rm \mu}$C/cm$^2$ and $P$ = -0.96 ${\rm \mu}$C/cm$^2$ are denoted by $P\uparrow$ and $P\downarrow$ respectively, the MAE dependency on the ferroelectric polarization for the In$_2$Se$_3$/Fe bridge-site adsorbed graphene monolayer is presented in Fig. 7.

%%%%%%%%%%%%%%%%Figure 7 phase diagram
\begin{figure*}[!ht]
\includegraphics[angle=0,width=0.7\textwidth]{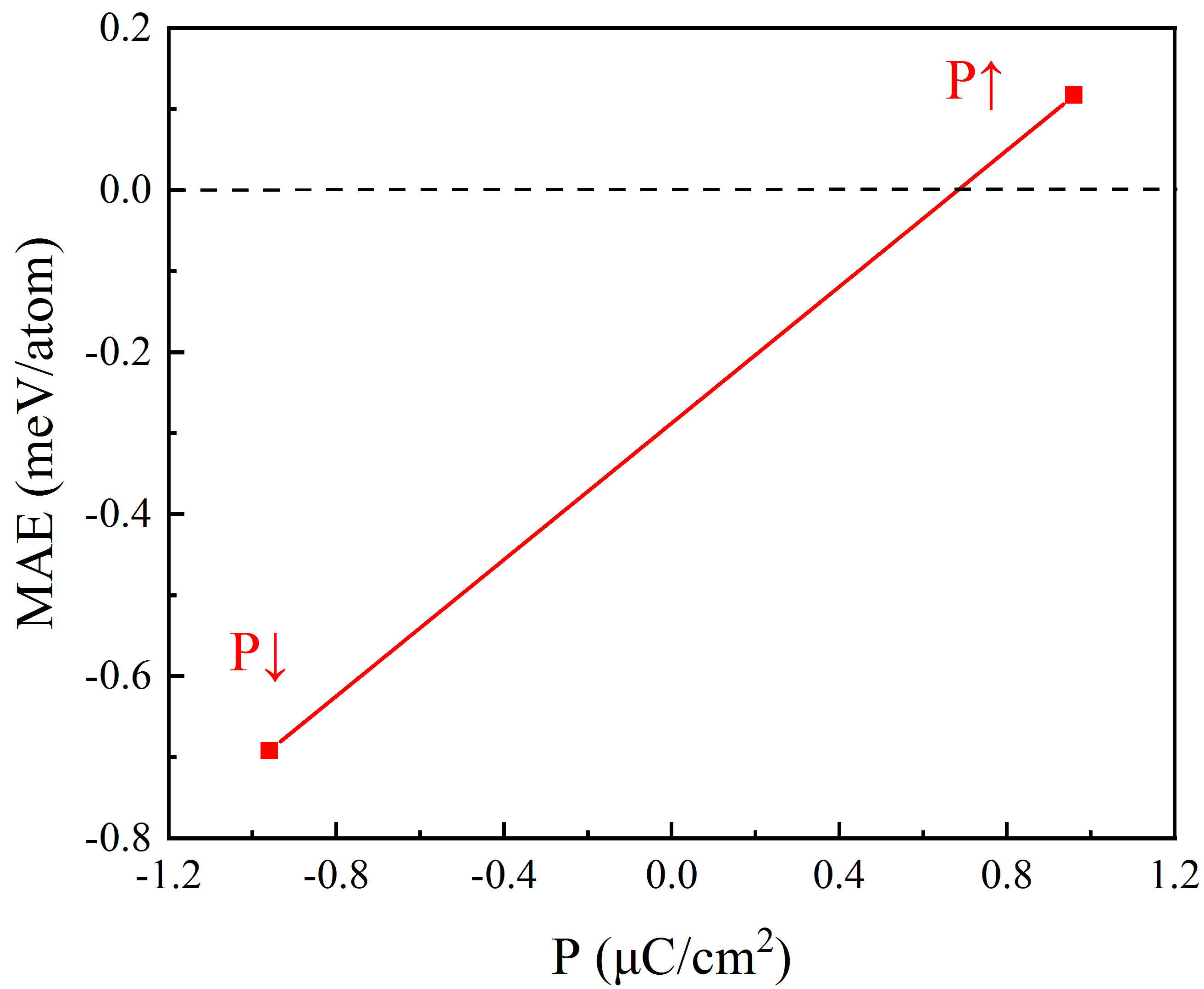}
\caption{\label{fig:7}  
The MAE dependency on the ferroelectric polarization for the In$_2$Se$_3$/Fe bridge-site adsorbed graphene monolayer.
}
\end{figure*}

To clarify the origin of MAE with spin states, qualitative analysis can be carried out based on second-order perturbation theory[31]:
% \begin{equation}
%     {\rm MAE} \simeq \xi^2
% \end{equation}

\begin{equation*}
\mathrm{MAE} \simeq \xi^{2} \sum_{\mathrm{o}, \mathrm{u}} \frac{\left|\left\langle\mathrm{o}\left|l_{z}\right| \mathrm{u}\right\rangle\right|^{2}-\left|\left\langle\mathrm{o}\left|l_{x}\right| \mathrm{u}\right\rangle\right|^{2}}{\varepsilon_{\mathrm{u}}-\varepsilon_{\mathrm{o}}} \tag{2}
\end{equation*}

Here, $\xi$ is the SOC constant, o and $u$ represent the eigenstates of occupied and unoccupied states, and $\varepsilon_{\mathrm{o}}$ and $\varepsilon_{\mathrm{u}}$ represent the eigenvalues of $o$ and $u$. For states with opposite and same spins, the MAEs can be further written as:

\begin{align*}
& \mathrm{MAE} \simeq-\xi^{2} \sum_{\mathrm{o}^{+}, \mathrm{u}^{-}} \frac{\left|\left\langle\mathrm{o}^{+}\left|l_{z}\right| \mathrm{u}^{-}\right\rangle\right|^{2}-\left|\left\langle\mathrm{o}^{+}\left|l_{x}\right| \mathrm{u}^{-}\right\rangle\right|^{2}}{\varepsilon_{u}^{-}-\varepsilon_{o}^{+}}  \tag{3}\\
& \mathrm{MAE} \simeq \xi^{2} \sum_{\mathrm{o}^{-}, \mathrm{u}^{-}} \frac{\left|\left\langle\mathrm{o}^{-}\left|l_{z}\right| \mathrm{u}^{-}\right\rangle\right|^{2}-\left|\left\langle\mathrm{o}^{-}\left|l_{x}\right| \mathrm{u}^{-}\right\rangle\right|^{2}}{\varepsilon_{u}^{-}-\varepsilon_{o}^{-}} \tag{4}
\end{align*}

As a result, elements of the matrix as well as the energy differences can affect the MAE jointly. Table 2 shows the differences of matrix elements 
$\left|\left\langle\mathrm{o}^{+}\left|l_{z}\right| \mathrm{u}^{-}\right\rangle\right|^{2}-\left|\left\langle\mathrm{o}^{+}\left|l_{x}\right| \mathrm{u}^{-}\right\rangle\right|^{2} \text { and }$
$\left|\left\langle\mathrm{o}^{-}\left|l_{z}\right| \mathrm{u}^{-}\right\rangle\right|^{2}-\left|\left\langle\mathrm{o}^{-}\left|l_{x}\right| \mathrm{u}^{-}\right\rangle\right|^{2} \quad$ for Eqs. (3) and (4), which are utilized to analyze the contribution of specific orbitals to MAE.

The magnetic quantum numbers $m= \pm 2, m= \pm 1$, and $m=0$ can represent the $d_{x^{2}-y^{2} / x y}, d_{x z y z}$ and $d_{z^{2}}$ of $3 d$-orbitals, respectively. The states with $|\Delta m|=0$ provide the positive contribution for the coupling in the same spins $(\uparrow \uparrow$, $\downarrow \downarrow)$ and the negative contribution for the coupling with different spins ( $\uparrow \downarrow, \downarrow \uparrow$ ); Whereas, the states with $|\Delta m|=1$ give the negative contribution for the coupling in the same spins ( $\uparrow \uparrow, \downarrow \downarrow$ ) and the positive contribution for the coupling in the different spins ( $\uparrow \downarrow, \downarrow \uparrow$ ).

%%%%%%%%%%%%%%%%Figure 8 phase diagram
\begin{figure*}[!ht]
\includegraphics[angle=0,width=0.7\textwidth]{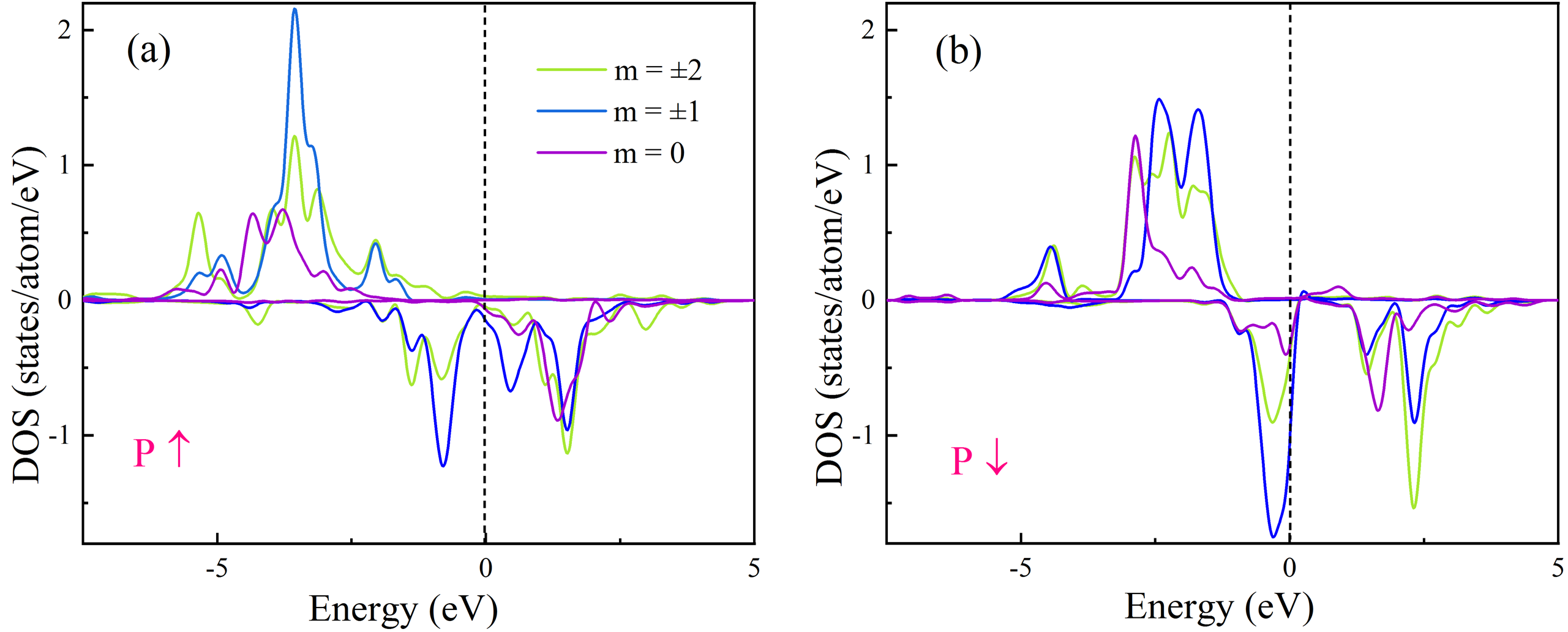}
\caption{\label{fig:8}  
The $3d$-orbital decomposed PDOS of the $\mathrm{Fe}$ atom at the $\mathrm{In}_{2} \mathrm{Se}_{3} / \mathrm{Fe}$ adsorbed graphene (bridge-site) heterojunctions for (a)  $P\uparrow$ and (b) $P\downarrow$. $m=0$, $m= \pm 1$, and $m= \pm 2$ are represented by purple line, blue line, and green line, respectively.
}
\end{figure*}

Fig. 8 presents the orbital-decomposed PDOS of Fe's $3 d$-orbitals for the $\mathrm{In}_{2} \mathrm{Se}_{3} / \mathrm{Fe}$ adsorbed-graphene (bridge-site) heterojunctions. Firstly, when the ferroelectric polarization of $\mathrm{In}_{2} \mathrm{Se}_{3}$ turns from upward to downward, the occupied $m= \pm 1 \downarrow$ states move towards the Fermi level while the unoccupied $m=0 \downarrow \quad$ states stand still, which diminishes the energy difference $\varepsilon_{u \sigma^{\prime}}-\varepsilon_{u \sigma}$. The orbitals $d_{y z}$ and $d_{z^{2}}$ with same spins can decrease the MAE through $\left\langle m= \pm 1 \downarrow\left|l_{x}\right| m=0 \downarrow\right\rangle$ for they yield a matrix element difference equalling -3 (Table 2). In addition, occupied $m= \pm 2 \uparrow$ states and unoccupied $m= \pm 2 \downarrow \quad$ states all move to the higher energy level, but the energy difference $\varepsilon_{u \sigma^{\prime}}-\varepsilon_{u \sigma}$ decreases. The orbitals $d_{x^{2}-y^{2}}$ and $d_{x y}$ with opposite spins can sharply diminish the MAE through $\left\langle m= \pm 2 \uparrow\left|l_{z}\right| m= \pm 2 \downarrow\right\rangle$ for they have a matrix element difference equalling -4 (Table 2). Consequently, the sum of the two contributions can reduce the MAE from 0.117 $\mathrm{meV}/$atom to -0.692 $\mathrm{meV}/$atom as the ferroelectric polarization of $\mathrm{In}_{2} \mathrm{Se}_{3}$ turns from upward to downward. It should be emphasized that the above analysis also applies to the decomposed PDOS obtained by HSE06 (see Fig. S10).

%%%%%%%%%%%%%%%%Figure 9 phase diagram
\begin{figure*}[!ht]
\includegraphics[angle=0,width=0.7\textwidth]{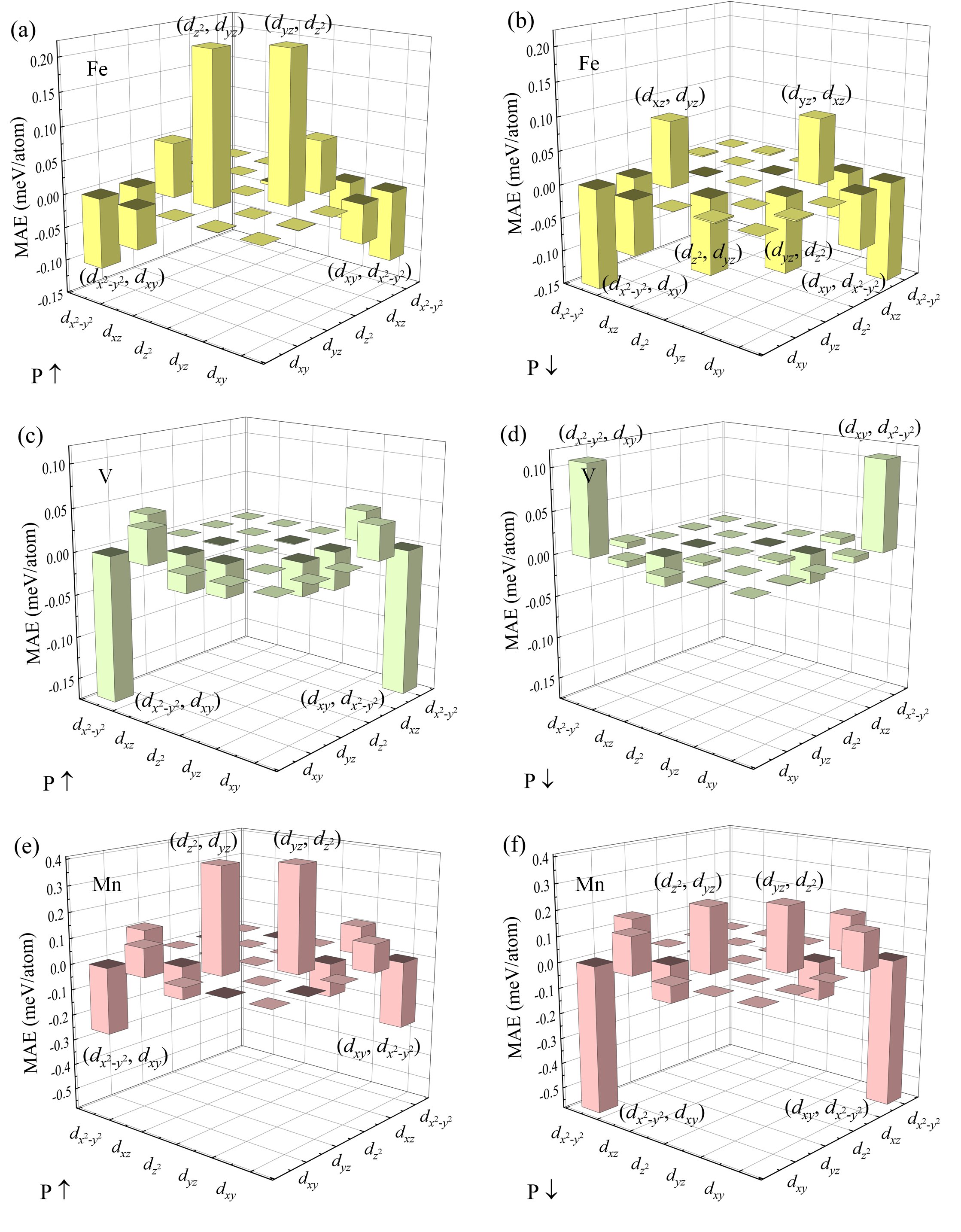}
\caption{\label{fig:9}  
The $3d$-orbital decomposed MAE of TM atom at the $\mathrm{In}_{2} \mathrm{Se}_{3} / \mathrm{TM}$ atom adsorbed graphene. (a) and (b) are of the $\mathrm{Fe}$ atoms, (c) and (d) are of the $\mathrm{V}$ atoms, and (e) and (f) are of the Mn atoms. The ferroelectric polarization direction of $\mathrm{In}_{2} \mathrm{Se}_{3}$ is upward in (a)(c)(e) and downward in (b)(d)(f).
}
\end{figure*}

To quantitatively proof the contribution of specific orbitals
to MAE, the $3 d$-orbital decomposed MAE of Fe in the upward and downward polarizations are shown in Fig. 9(a) and Fig. 9(b). It's obvious that for upward polarization in Fig. 9(a), elements of the matrix between $d_{z^{2}}$ and $d_{y z}$ yield large positive values for MAE which are more dominant than those showing negative values, thus leading to a positive MAE of 0.117 $\mathrm{meV}/$atom; For downward polarization in Fig. 9(b), elements of the matrix between different orbitals except for $d_{x z}$ and $d_{y z}$ all show negative values, therefore the MAE has a very large negative value of -0.692 $\mathrm{meV}/$atom. The $3 d$-orbital-decomposed MAEs of the interface $\mathrm{V}$ and $\mathrm{Mn}$ atoms are calculated and the corresponding results are shown in Fig. 9(c-d) and (e-f), respectively. The hybrid matrix element $d_{x^{2}-y^{2}}$ and $d_{x y}$ is crucial for the $\mathrm{V}$ atoms, which provides negative (positive) contributions to $\mathrm{P} \uparrow(\mathrm{P} \downarrow)$, thus leading to a MAE of -0.403 $\mathrm{meV}/$atom (0.516 $\mathrm{meV}/$atom), respectively (Fig. 9(c-d)). For Mn atoms in Fig. 9(e-f), elements of the matrix $d_{z^{2}}$ and $d_{y z}$ yield positive contributions for MAE, while those between $d_{x^{2}-y^{2}}$ and $d_{x y}$ yield negative contributions for MAE; Since the former contributions are more dominant for $\mathrm{P} \uparrow$ but the latter ones are more dominant for $\mathrm{P} \downarrow$, the MAE becomes 0.255 $\mathrm{meV}/$atom for $\mathrm{P} \uparrow$ and -0.364 $\mathrm{meV}/$atom for $\mathrm{P} \downarrow$, respectively.

Fig. 10 presents the plane-averaged charge density difference $\Delta \rho(\mathrm{z})$ with $\mathrm{P} \uparrow$ and $\mathrm{P} \downarrow$. The results are in line with expectations that upward polarization causes electron depletion near TM atoms (see Fig. 10(a)(c)(e)), while downward polarization causes electron accumulation near TM atoms (see Fig. 10(b)(d)(f)). The charge transfers can bring changes in the amounts and occupations of the majority and minority spin electrons in TM atoms, thus affecting the magnetic moments and MAEs.

%%%%%%%%%%%%%%%%Figure 10 phase diagram
\begin{figure*}[!ht]
\includegraphics[angle=0,width=0.7\textwidth]{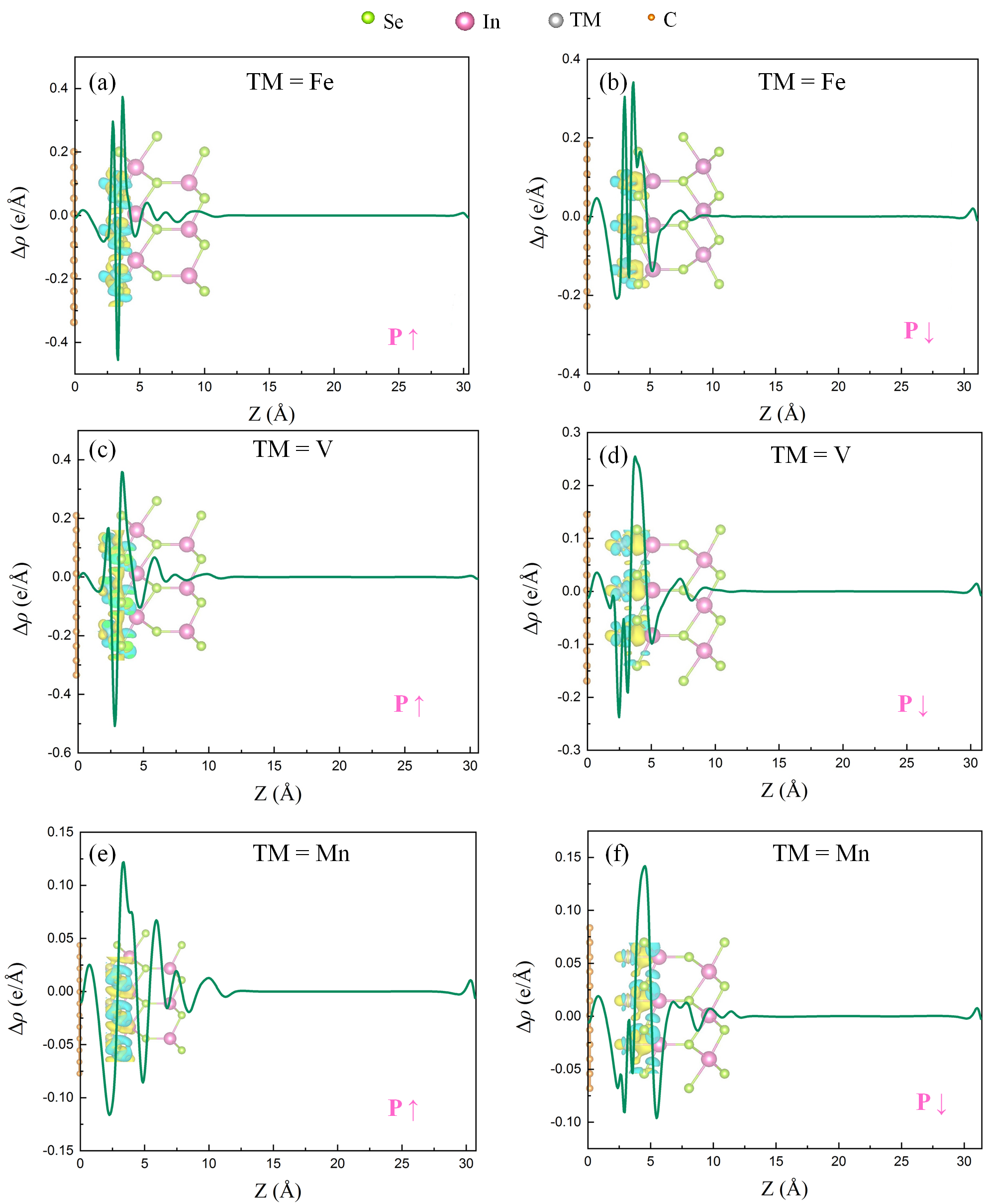}
\caption{\label{fig:10}  
Plane-averaged charge density difference $\Delta \rho(\mathrm{z})$ with upward ferroelectric polarization in (a)(c)(e) and downward ferroelectric polarization in (b)(d)(f): (a) and (b) are for $\mathrm{Fe}$ atoms, (c) and (d) are for $\mathrm{V}$ atoms, and (e) and (f) are for Mn atoms. Here, blue and yellow regions refer to electron's depletion and accumulation, respectively.
}
\end{figure*}

\begin{center}
\begin{tabular}{lcccccccccc}
\hline\hline
% u$^-$ & \multicolumn{4}{c}{$o^{+}$} & &  &  \multicolumn{1}{c}{$o^{-}$}  &  &  &  \\ 
% \cline{2-6} \cline{7-11}
\multirow{2}{*}{u$^-$} &  & \multicolumn{4}{c}{$o^{+}$} &  & &  \multicolumn{1}{c}{$o^{-}$} &  &  \\
\cline{2-6} \cline{7-11}
 & $d_{x y}$ & $d_{y z}$ & $d_{z^{2}}$ & $d_{x z}$ & $d_{x^{2}-y^{2}}$ & $d_{x y}$ & $d_{y z}$ & $d_{z^{2}}$ & $d_{x z}$ & $d_{x^{2}-y^{2}}$ \\
$d_{x y}$ & 0 & 0 & 0 & 1 & -4 & 0 & 0 & 0 & -1 & 4 \\
$d_{y z}$ & 0 & 0 & 3 & -1 & 1 & 0 & 0 & -3 & 1 & -1 \\
$d_{z^{2}}$ & 0 & 3 & 0 & 0 & 0 & 0 & -3 & 0 & 0 & 0 \\
$d_{x z}$ & 1 & -1 & 0 & 0 & 0 & -1 & 1 & 0 & 0 & 0 \\
$d_{x^{2}-y^{2}}$ & -4 & 1 & 0 & 0 & 0 & 4 & -1 & 0 & 0 & 0 \\
\hline\hline
\end{tabular}
\end{center}

\section{Conclusions}
In summary, through first-principles calculations, we have demonstrated the manipulation of switching the magnetization orientations of magnetized graphene monolayers by controlling their adjacent ferroelectric polarizations in 2D van der Waals heterostructures. Specifically speaking, the easy magnetization axis of the V, Mn, and Fe adsorbed graphene monolayers can be flipped by the ferroelectric polarizations of adjacent $\mathrm{In}_{2} \mathrm{Se}_{3}$ layers. For instance, when the $\mathrm{Fe}$ atom is adsorbed on the graphene monolayer at the bridge site, the direction of magnetization undergoes a flip from along [001] (MAE =0.117 $\mathrm{meV}/$atom) to along [100] (MAE =-0.692 $\mathrm{meV}/$atom) if the ferroelectric polarization of $\mathrm{In}_{2} \mathrm{Se}_{3}$ turns from upward to downward. It is mainly due to that the MAE source from the matrix element between $d_{z^{2}}$ and $d_{y z}$ changes from positive to negative. This work demonstrates a prospective path to realizing the ferroelectric-control of magnetism in two-dimensional ferroelectric/ferromagnetic heterostructures and may further stimulate the theoretical and experimental exploration in studying magnetized graphene derivatives and other 2D spintronic materials.

\section{Data availability}
\section*{Supplementary material}
See the supplementary material for the convergence tests of total energy and MAE (Fig. S1-S6), the dependency between the value of MAE and $U_{\text {eff }}$ (Fig. S7-S9), and the $3 d$-orbital decomposed PDOS obtained by HSE06 (Fig. S10).

\section*{Acknowledgments}
This study is supported by Scientific Research Program Funded by Education Department of Shaanxi Provincial Government (Program No. 21JK0699) and Natural Science Basic Research Program of Shaanxi (Program No. 2024JC-YBQN-0038).

\section*{Declaration of Competing Interest}
There are no conflicts that need to be declared.

\section{Acknowledgments}

\section{Author contributions}
Y.-W.F. conceived the study and planned the research. Y.-W.F. and T.-M.L. supervised the project.
R.-Q.W. performed and analyzed the theoretical calculations with assistance from Y.-W.F.. 
The manuscript was drafted by R.-Q.W. and further revised by all the authors. 

\section{Competing interests}
The authors declare no competing interests.

\section{Additional information}
Materials \& Correspondence should be addressed to Y.-W.F.

\clearpage
\newpage
\appendix

\setcounter{figure}{0} % Reset figure counter to 0
\renewcommand\thefigure{S\arabic{figure}} % Redefine figure numbering format as Fig. S1, S2, etc.
\renewcommand{\theHfigure}{S\arabic{figure}}% Hyperref figure hyperlink hook %reference: https://tex.stackexchange.com/questions/519879/the-reference-of-the-first-figure-in-the-appendix-leads-to-the-first-figure-in-t

\setcounter{table}{0} % Reset figure counter to 0
\renewcommand\thetable{S\arabic{table}} % Redefine figure numbering format as Fig. S1, S2, etc.

% \section{Properties of lowest-enthalpy ternary state at 1 GPa: ${P2_1/m}$ LuH$_2$N }
% Fig.~\ref{fig:2fu_LuH2N_389_phonon_supercell222} shows the phonon dispersion and element projected phonon DOS of ${P2_1/m}$ LuH$_2$N (ID: 2fu\_LuH2N\_389) calculated by VASP and Phonopy. Fig.~\ref{fig:2fu_LuH2N_389_e_dos} displays the calculated electronic DOS.
% %%%%%%%%%%%%%%%%%%%%%
% %%%%%%%%%%%%%%%%Figure phonon properties from supercell methods for 2fu_LuH2N_389
% \begin{figure}[ht]
% \includegraphics[angle=0,width=0.99\textwidth]{2fu_LuH2N_389-phonon-supercell222}
% \caption{\label{fig:2fu_LuH2N_389_phonon_supercell222}  
% \textbf{The phonon dispersion and element projected phonon DOS of ${P2_1/m}$ LuH$_2$N. } 
% }
% \end{figure}

[1] H.-X. Cheng, J. Zhou, C. Wang, W. Ji, Y.-N. Zhang, Phys. Rev. B 104, 064443 (2021).

[2] B. S. Yang, B. Shao, J. F. Wang, Y. Li, C. Y. Yam, S. B. Zhang, B. Huang, Phys. Rev. B 103, L201405 (2021).

[3] Y. Lu, R. X. Fei, X. B. Lu, L. H. Zhu, L. Wang, ACS Appl. Mater. Interfaces 12(5), 6243-6249 (2020).

[4] M. H. Liu, L. Zhang, J. X. Liu, T. L. Wan, A. J. Du, Y. T. Gu, L. Z. Kou, ACS Appl. Electron. Mater. 5, 920-927 (2023).

[5] W. Sun, W. X. Wang, D. Chen, Z. X. Cheng, Y. X. Wang, Nanoscale 11, 9931-9936 (2019).

[6] Y. P. Wang, X. G. Xu, X. Zhao, W. X. Ji, Q. Cao, S. S. Li, Y. L. Li, npj Comput. Mater. 8, 218 (2022).

[7] R. Li, J. W. Jiang, W. B. Mi, H. L. Bai, Appl. Phys. Lett. 120, 162401 (2022).

[8] B. X. Zhai, R. Q. Cheng, W. Yao, L. Yin, C. H. Shen, C. X. Xia, J. He, Phys. Rev. B 103, 214114 (2022).

[9] W. R. Liu, X.-J. Dong, Y.-Z. Lv, W.-X. Ji, Q. Cao, P.-J. Wang, F. Li, C.-W. Zhang, Nanoscale 14, 3632 (2022).

[10] A. Ilyas, S. L Xiang, M. G. Chen, M. Y. Khan, H. Bai, P. He, Y. H. Lu, R. R. Deng, Nanoscale 13, 1069 (2021).

[11] D. Y. Jin, W. Qiao, X. Y. Xu, W. B. Mi, S. M. Yan, D. H. Wang, Appl.Surf.Sci. 580, 152311 (2022).

[12] W. J. Ding, J. B. Zhu, Z. Wang, Y. F. Gao, D. Xiao, Y. Gu, Z. Y. Zhang, and W. G. Zhu, Nat. Commun. 8, 14956 (2017). 

[13] S. Y. Wan, Y. Li, W. Li, X. Y. Mao, C. Wang, C. Chen, J. Y. Dong, A. M. Nie, J. Y. Xiang, Z. Y. Liu, W. G. Zhu, H. L. Zeng, Adv. Funct. Mater. 29, 1808606 (2019).

[14] C. Gong, L. Li, Z. Li, H. Ji, A. Stern, Y. Xia, T. Cao, W. Bao, C. Wang, Y. Wang, Z. Q. Qiu, R. J. Cava, S. G. Louie, J. Xia and X. Zhang, Nature 546, 265–269 (2017).

[15] B. Huang, G. Clark, E. Navarro-Moratalla, D. R. Klein, R. Cheng, K. L. Seyler, D. Zhong, E. Schmidgall, M. A. McGuire, D. H. Cobden, W. Yao, D. Xiao, Nature 546, 270–273 (2017).

[16] H. L. Zhuang, P. R. C. Kent and R. G. Hennig, Phys. Rev. B 93, 134407 (2016).

[17] H. Sun, S.-S. Li, W.-X. Ji, C.-W. Zhang, Phys. Rev. B 105, 195112 (2022).

[18] B. Wu, Y.-L. Song, W.-X. Ji, P.-J. Wang, S.-F. Zhang, C.-W. Zhang, Phys. Rev. B 107, 214419 (2023).

[19] S.-J. Zhang, C.-W. Zhang, S.-F. Zhang, W.-X. Ji, P. Li, P.-J. Wang, S.-S. Li, S.-S. Yan, Phys. Rev. B 96, 205433 (2017).

[20] X. Chen, Y. H. Liu, B. Sanyal, J. Phys. Chem. C 124(7), 4270-4278 (2020).

[21] Z. Y. Guan, S. Ni, S. L. Hu, ACS Omega. 5, 5900-5910 (2020).

[22] S. Stankovich, D. A. Dikin, G. H. B. Dommett, K. M. Kohlhaas, E. J. Zimney, E. A. Stach, R. D. Piner, S. T. Nguyen, R. S. Ruoff, Nature 442, 282–286 (2006).

[23] R.-Q. Wang, Y.-W. Fang, T.-M. Lei, J. Magn. Magn. Mater. 565, 170297 (2023).

[24] S. L. Dudarev, G. A. Botton, S. Y. Savrasov, C. J. Humphreys, A. P. Sutton, Phys. Rev. B 57, 1505 (1998).

[25] I. V. Solovyev, P. H. Dederichs, V. I. Anisimov, Phys. Rev. B. 50 (23), 16861–16871 (1994).

[26] S. Grimme, J. Antony, S. Ehrlich, and S. Krieg, J. Chem. Phys. 132, 154104 (2010).

[27] J. Paier, M. Marsman, K. Hummer, G. Kresse, I. C. Gerber, and
J. G. Ángyán. J. Chem. Phys. 124, 154709 (2006).

[28] A. Banerjea, J. R. Smith, Phys. Rev. B 37, 6632–6645 (1988).

[29] D. Vanderbilt, R. D. King-Smith, Phys. Rev. B 48, 4442 (1993).

[30] X. X. Jiang, Y. X. Feng, K. Q. Chen, L. M. Tang, J. Phys. Condens. Matter 32, 105501 (2020).

[31] D.-S. Wang, R. Q. Wu, A. J. Freeman, Phys. Rev. B 47, 14932 (1993).

% \bibliography{polar_metal}
\end{document}